\documentclass[12pt,a4paper]{article}
\usepackage{latexsym,amsmath,amssymb}
\date{}
\numberwithin{equation}{section}

\begin{document}
\title{Krein structure of supersymmetry}
\author{Florin Constantinescu\\ Fachbereich Mathematik \\ Johann Wolfgang Goethe-Universit\"at Frankfurt\\ Robert-Mayer-Strasse 10\\ D 60054
Frankfurt am Main, Germany}
\maketitle

\begin{abstract}
We present supersymmetric positive definite scalar products together with  natural Krein structures of supersymmetries.
\end{abstract}

\section{Introduction}

Supersymmetries generalize the notion of a Lie algebra to include
algebraic systems whose defining relations involve commutators as well
as anticommutators. Denoting schematically by $ Q_{\alpha },\bar Q_{\dot
  \alpha } $ the odd (anticommuting) generators of supersymmetry, physical considerations require that (see \cite {WB}) the operators $ Q_\alpha  ,\bar Q_{\dot \alpha }=(Q_\alpha)^+ $ act in a bona fide Hilbert space of states with positive definite metric. From the commutation relation
\begin{equation}\nonumber
\{Q_{\alpha },\bar Q_{\dot \alpha }\}=2\sigma_{\alpha \dot \alpha }^\mu P_\mu 
\end{equation}
it follows that for any state $ \Phi $ 
\begin{equation}\nonumber
(\Phi,\{Q_{\alpha },\bar Q_{\dot \alpha }\}\Phi)=2\sigma_{\alpha \dot \alpha }^\mu (\Phi ,P_\mu \Phi )
\end{equation}
Summing over $\alpha ,\dot \alpha =1,2 $ and using tr $ \sigma ^l=2\delta^{l0}, l=0,1,2,3 $ where $\sigma^l $ are the Pauli matrices yields
\begin{equation}\nonumber
4(\Phi, P_0\Phi )>0
\end{equation}
i.e. in a supersymmetric theory the energy $ H=P_0 $ is always positive (for notations and conventions see next section). This beautiful argument doesn't require any detailed knowledge of the Hilbert space. On the other hand in quantum mechanics and quantum field theory (free fields or tentatively perturbation theory formulated in the Hilbert space of the non-interacting theory) Hilbert spaces do have realizations as some $ L^2$-spaces of functions \cite{J,StreW} and \cite{EG}. In this paper we present invariant positive definite scalar products in spaces of supersymmetric functions and show that the corresponding Hilbert spaces appear in conjunction with  natural Krein structures which seem to be inherent to superymmetries.

\section{Notations and conventions}
We use the notation and conventions of \cite{WB} with the only difference that $ \sigma ^0, \bar \sigma ^0 $
are the identity instead of minus identity (our notations coincide with \cite {S}). In particular the Minkowski metric is $(-1,1,1,1)$.\\
We use for the time being formal projections
\begin{gather} \nonumber
P_1=P_a=\frac {1}{16\square } D^2 \bar D^2 \\ \nonumber
P_2=P_c=\frac {1}{16\square } \bar D^2 D^2  \\ \nonumber
P_T=-\frac{1}{8\square }D\bar D^2 D=-\frac{1}{8}\bar D D^2\bar D \nonumber
\end{gather}
with properties $P_i^2=P_i,P_iP_j=0,i\neq j,i=c,a,T $ and $ P_c+P_a+P_T=1 $. Here $D,\bar D$ with spinor component $D_{\alpha}, \bar D_{\dot \alpha} $ are the covariant (and invariant) derivatives given by
\begin{gather}
D_{\alpha }=\partial_{\alpha } +i\sigma_{\alpha \dot \alpha }^l\bar \theta^{\dot \alpha }\partial _l \\
\bar D_{\dot \alpha }=-\partial_{\dot \alpha } -i\theta^{\alpha }\sigma _{\alpha \dot \alpha }^l \partial _l 
\end{gather}
where $\sigma^l ,l=0,1,2,3 $ are the Pauli matrices.
At the first glance the d'alembertian in the denominator of the $P$-projections may produce some problems. It can be shown that this is not the case. In the ´"massles" case this requires some technical details which we do not enter here.

\section{Supersymmetric functions and distributions}

The operators in the preceding section act on supersymmetric functions in the variable $z=(x,\theta ,\bar \theta ) $:
\begin{gather} \nonumber
X(z)=X(x,\theta ,\bar \theta )= \\ \nonumber
=f(x)+\theta \varphi (x) +\bar \theta \bar \chi (x) +\theta ^2m(x)+\bar \theta^2n(x)+ \\
\theta \sigma^l\bar \theta v_l(x)+\theta^2\bar \theta \bar \lambda(x)+\bar \theta^2\theta \psi (x)+ \bar \theta^2 \theta^2d(x)
\end{gather}
where the coefficients are regular functions of $x=(x^0,\bar x) $ of fast decrease at infinity. Eventually we will consider also functions of several variables $z_1,z_2\ldots $. It can be shown that the conditions $\bar D_{\dot \alpha }X=0,\dot \alpha =1,2 $ are equivalent to $ P_cX=X $ and define chiral functions of the form
\begin{gather}\nonumber
X_c(z)=f(x)+\sqrt 2\theta \varphi (x) +\theta ^2m(x)+i\theta \sigma^l \bar \theta \partial_l f(x)+\\ 
+\frac{i}{\sqrt 2}\theta ^2\partial_l \varphi (x)\sigma ^l\bar \theta +\frac{1}{4}\theta ^2\bar \theta^2 \square f(x) 
\end{gather}
Similarly the conditions $D_\alpha X=0, \alpha =1,2 $
and $P_aX=X $ are equivalent and define antichiral functions
\begin{gather}\nonumber
X_a(z)=f(x)+\sqrt 2\bar \theta \bar \chi (x) +\bar \theta ^2n(x)+i\theta \sigma^l \bar \theta \partial_l f(x)+\\ 
+\frac{i}{\sqrt 2}\bar \theta ^2\theta \sigma^l\partial_l \bar \chi(x) +\frac{1}{4}\theta ^2\bar \theta^2 \square f(x) 
\end{gather}
The conjugate of a chiral is an antichiral function and the other way round. \\
Finally the conditions $D^2X=\bar D^2X=0$ and $P_TX=X$ are equivalent. They define the transverse functions of the form
\begin{gather} \notag
X_T(z)=f(x)+\sqrt 2\theta \varphi (x) +\sqrt 2\bar \theta \bar \chi (x) + \theta \sigma^l\bar \theta v_l(x)- \\ 
-\frac{i}{\sqrt 2}\theta ^2\partial_l \varphi (x) \sigma^l \bar \theta -\frac{i}{\sqrt 2} \bar \theta^2 \theta \sigma^l \partial_l \bar \chi (x)-\frac{1}{4}\theta^2 \bar \theta^2 \square f(x)
\end{gather}
with the supplimentary condition div $v(x)=0 $.
Whereas the proof of (3.2),(3.3) is rather standard \cite{WB} the proof
of (3.4) requires more computational work and can be found in
\cite{S}. It makes use of the characterization of the transverse
functions $X$ by $D^2X=\bar D^2X=0$. Note the similarity between
(3.2),(3.3) on one side and (3.4) on the other side. But there is a
difference concerning the signs in the derivatives of $f, \varphi $ and
$\bar \chi $ which will have major consequences in the following.\\
Together with (test)functions (3.1) we consider superdistributions of the same form (3.1) where the coefficients are ordinary distributions.\\
Now consider the following set of functions (or distributions) $K_i(z_1,z_2),i=0,c,a,T$ of two variables $z_1,z_2$: 
\begin{equation}
K_0(z_1-z_2)=\delta(\theta_1 -\theta_2)\delta(\bar \theta_1 -\bar \theta_2 )F(x_1-x_2)
\end{equation}
and
\begin{gather}
K_c(z_1,z_2)=P_cK_0(z_1-z_2) \\
K_a(z_1,z_2)=P_aK_0(z_1-z_2) \\
K_T(z_1,z_2)=-P_TK_0(z_1-z_2)
\end{gather}
where $\delta(\theta _1-\theta_2)=(\theta_1-\theta _2)^2, \delta(\bar \theta _1-\bar \theta_2)=(\bar \theta_1-\bar \theta _2)^2 $ and $F(x)$ is a positive definite Poincare invariant function (distribution) concentrated in the forward light cone. By convention it is understood that the derivatives in $P_iK_0(z_1-z_2)$ act on the first variable. For definiteness the reader can keep in mind as an example for (3.5) a kernel $K_{0m}$ with the function (distribution) $F$ given by the  Pauli-Jordan function $F(x)=-iD_m^+(x)$ of mass $m\geqslant 0$:
\begin{equation}
D_m^+(x)=\frac{i}{(2\pi)^3}\int \theta(p_0)\delta(p^2-m^2)e^{ipx}dp
\end{equation}
In (3.9) we use standard conventions (see for instance \cite{J,StreW}) but in the metric $(-1,1,1,1)$.When we want to specify the mass in $K_0$ we write $K_{00}$ for the massless and $ K_{0m}$ for the massive case.\\
In the next section the kernels $K_i(z_1,z_2)$ and related ones will be used in order to construct supersymmetric scalar products. Note that these functions (with the exception of $K_0 $) will depend on $\theta-$ variables not only through the differences. Note also the surprising minus sign in front of transverse projection (3.8) which is somewhat at odd with the relation $P_c+P_a+P_T=1$ and will be responsible for the generic Krein structure of the supersymmetry. This will be the main result of the paper.

\section{Supersymmetric scalar products}

With the help of the kernels $K_i$ we define inner products of the
form
\begin{equation}
(X,Y)=\int d^8z_1d^8z_2\overline {X(z_1)}K_i(z_1,z_2)Y(z_2)
\end{equation}
where the bar on the rhs means conjugation (including Grassmann) and $d^8z=d^4xd^2\theta d^2\bar \theta $.
They are compatible with the conjugation i.e.
$\overline {(X,Y)}=(Y,X)$ (hermiticity). We will prove that for $i=c,a,T$ the above inner products are positive (more precisely non-negative) definite, i.e. after factorization of kernels and completion they define (strict) positive definite scalar products. In fact in our terminology we will make no difference between (strict) positivity and non-negativity; factorization followed by completion is always understood. As far as the case $i=0$ is concerned, the kernel $K_0$ defines a positive scalar product $(X,Y)$ only if $X,Y$ are restricted to linear combinations of chiral and antichiral functions. It is not of special physical interest but it serves as a key point for the proofs of the positivity for other scalar products which are more interesting from the physical point of view.\\
Let us give now some details concerning proofs of positivity for our scalar products. We start with the particular case of the kernel $K_0 (z_1-z_2)$. For chiral $X=X_c$ and $Y=Y_c$ we have
\begin{equation}
(X_c,Y_c)=\int d^8z_1d^8z_2\overline {X_c(z_1)}K_0(z_1-z_2)Y_c(z_2)
\end{equation}
By integrating one set of $\theta $-variables the rhs in (4.2) can be written as
\[ \int d^4x_1d^4x_2d^2 \theta d^2 \bar \theta \overline {X_c(x_1,\theta ,\bar \theta )}F(x_1-x_2)Y_c(x_2,\theta, \bar \theta )        \]
For $Y_c=X_c$ the positivity follows from a computation in momentum space using the general form of chiral and antichiral function in (3.2),(3.3) and remarking that $\sigma p$ and $-p^2$ are positive when restricted to the forward light cone. The same is true in the antichiral sector. On the other hand integrals of chiral (antichiral) against antichiral (chiral) function vanish. This proves the result together with a direct sum decomposition of the chiral/antichiral sectors. The reader can convince himself that this decomposition does not extend to include the transverse sector as this might be suggested by $P_c+P_a+P_T=1$.\\
We can pass now to the cases $i=c,a$ by using the projection property $P_i^2=P_i, i=c,a$ and integration by parts in superspace (see for instance \cite{S}). Indeed the integral
\[\int d^8z_1d^8z_2 \overline {X(z_1)}P_c K_0(z_1-z_2)X(z_2)=\int d^8z_1d^8z_2  \overline {X(z_1)}P_c^2 K_0(z_1-z_2)X(z_2)\]
can be written by partial integration as
\begin{gather}\notag
\int d^8z_1d^8z_2 \overline {P_c X(z_1)}K_0(z_1-z_2)P_c X(z_2)= \\
=\int d^4x_1d^4x_2d^2\theta d^2\bar \theta \overline {P_c X(x_1,\theta ,\bar \theta)}F(x_1-x_2)P_cX(x_2,\theta ,\bar \theta)
\end{gather}
We have a similar relation for the antichiral case. In this way the scalar products for the case $i=c,a$ are projected down to the chiral/antichiral sectors and we can apply the previous result. \\
Even more interesting is the inner product generated by the kernel $-P_TK_0(z_1,z_2)$. Note the minus sign in front of $P_T$. It appears in connection whith the supersymmetric massive or massless vector field (see also the next section). To prove positivity in this case we also use the projection property of $P_T$ as above and then apply integration by parts. As a result we have to compute 
\begin{gather}\notag
\int d^8z_1d^8z_2 \overline {P_T X(z_1)}K_0(z_1-z_2)P_TX(z_2)= \\
=\int d^4x_1d^4x_2  \overline {P_T X(x_1,\theta ,\bar \theta )}F(x_1-x_2)P_TX(x_2,\theta ,\bar \theta )
\end{gather}
In order to obtain the positivity of (4.1) for $i=T$ we have to apply in
momentum space beside the previous argument regarding the signs of
$\sigma p$ and $-p^2$ a further argument concerning the positivity of
the Minkowski scalar product $\bar v_lv^l$ for functions satisfying the
property div$v=0$. In momentum space the condition div$v(x)=0$
translates into the orthogonality of the vector function $v=(v^l(p))$ to
the momentum vector $p=(p_l)$ representing the independent
variable. This is necessary in order to obtain a positive contribution
to the scalar product from the vector part $v$ of (3.1). In physics the
vector field $v$ is real but the argument works even for complex
functions. This argument is not new; it was used long time ago \cite{WG,SW} in connection with a rigorous Gupta-Bleuler formalism in electrodynamics. What seems to be new in supersymmetry is the fact that the extra condition div$v=0$ which has to be imposed ad hoc on test functions in quantum electrodynamics \cite{SW} comes here for granted from supersymmetric transversality (both in the massive as well in the massless case). In particular  it is possible to write down (even strict positive definite) Fock space representations of the free (massive) supersymmetric vector field (the supersymmetric generalization of the Proca field) along the lines of \cite{SW}. The positive mass is incompatible with gauge invariance. The scalar product in the massive case above (i.e. the scalar product generated with the help of $-P_T$) corresponds to the unitary gauge. The reader can convince himself that other cases, like for instance the massive vector field in the Feynman gauge, do not have Fock space representations compatible with positivity because the two point function contains chiral and antichiral contributions with a wrong sign. \\
In the more interesting gauge invariant massless case using the scalar
product generated with the help of $-P_T$ we can write down Fock space
representations of the vector field in the Wess-Zumino gauge and
arbitrary usual gauge of the vector contribution $v$. As in the
classical case of the electromagnetic field \cite{SW} the divergence
condition on the vector field div$v=0$, which is now a consequence of
supersymmetry, enforces positivity (in fact only non-negativity which
has to be turned into positivity by the same kernel factorization as in
\cite{SW}). Such representations even do not require a gauge fixing
(like the Wess-Zumino one) if we project in the test function space on
the transversal sector (cf. with the transversal projections $\Pi_{+,-}$
in (2.31),(2.32) of \cite{SW}) and are at the same time supersymmetric and gauge invariant. It would be interesting to investigate positive definite Fock space representations using free ghosts \cite{Wei} (including the pure super-Yang-Mills fields in which case several independent copies of the abelian case are needed \cite{Wei}; see also the remark about ghosts at the end of this section). \\ 
The considerations above show that the generic scalar product of supersymmetries is
\begin{equation}
(X,Y)=\int d^8z_1d^8z_2 \overline {X(z_1)}(P_c+P_a-P_T)K_0(z_1-z_2)Y(z_2)
\end{equation} \\
for arbitrary $X,Y$ of the form (3.1).
At the end of this section let us remark that there is still another scalar product which projects down to the chiral/antichiral sectors and which is interesting in supersymmetris. Restricting the $x-$part of the kernel to the Pauli-Jordan function, the matrix supersymmetric kernel of this scalar product is given by 
\begin{gather}
K_{WZ}(z_1,z_2)=\begin{pmatrix} \frac{1}{16}D^2\bar D^2 & \frac{m}{4}D^2 \\ \frac{m}{4}\bar D^2 & \frac{1}{16}\bar D^2D^2 \end{pmatrix}K_0(z_1-z_2)
\end{gather}
where $m>0$. We recognize on the diagonal (up to factors) the antichiral and chiral projections. The above kernel defines a scalar product in the set of two component supersymmetric functions. In fact it projects down to the chiral/antichiral sectors. The proof is again by using projection properties and partial integration in superspace. It gives the two point function of the (free) Wess-Zumino model. Being the two point function of a free field it can be used to give (and prove) explicit representations in the supersymmetric Fock-space (as bona fide Hilbert space), Wick ordering, Wick theorem and last but not least to set up the supersymmetric Epstein-Glaser \cite{EG} renormalization method for the $\phi^3$-Wess-Zumino model. Let us remark that the Fock spase of the (free) Wess-Zumino model is always symmetric in spite of the fact that the multiplet in question contains Majorana fields as well. On the contrary one expects that the corresponding chiral/antichiral ghosts are always antisymmetric. We will come back to these questions elsewhere.

\section{The Krein structure of supersymmetries}

In this section we present a generic Krein structure of supersymmetries. Let $V$ be an inner product space with inner product $<.,.>$ and $\omega $ an operator on $V$ with $\omega^2=1$. If $(\phi,\psi)=<\phi,\omega \psi> ; \phi ,\psi \in V$ is a (positive definite) scalar product on $V$ than we say that $V$ has a Krein structure. By completing in the scalar product (.,.) we obtain an associated Hilbert space structure (if $(.,.)$ has zero vectors we have in addition to factorize them before completing). Krein structures naturally appear in gauge theories (including the well understood case of electrodynamics) (see for instance the book \cite{Stro}).\\
The simplest supersymmetric Krein structure which emerges from the considerations of the preceding section is given by
\begin{equation}
<X,Y>=\int d^8z_1d^8z_2\overline {X(z_1)}K_0(z_1-z_2)Y(z_2)
\end{equation}
in the notations
$X=(P_c+P_a+P_T)X=X_c+X_a+X_T$ and subsequent identifications
$X=\begin{pmatrix}X_c \\X_a \\X_T \end{pmatrix}, K_0(z)=K_0(z)I_3 $ where $I_3$ is the 3x3 identity matrix. Now let
\begin{equation}
(X,Y)=<X,\omega Y>
\end{equation}
with 
\begin{equation}\nonumber
\omega=\begin{pmatrix}1& 0 &0 \\0 &1& 0 \\0& 0 & -1 \end{pmatrix}
\end{equation}
Certainly $(.,.)$ is positive definite being a transcription of (4.5).\\
A generalization of (5.1) together with its antichiral part is obtained as follows:
Let
\begin{equation}\nonumber
\mathcal {X}=\begin{pmatrix} X_1 \\ X_2 \\ X_3 \end{pmatrix}
\end{equation}
where $X_i,i=1,2,3$ are of the form (3.1) and introduce
\begin{equation}\nonumber 
\mathcal {K}(z_1,z_2)=\begin{pmatrix}P_c&0&0\\0&P_a&0\\0&0&P_T \end{pmatrix}K_0(z_1-z_2) 
\end{equation}
Note that $\mathcal {K}(z_1,z_2)$ does not depend only of the difference $z_1-z_2$.
The inner product in the three-component supersymmetric function space is
\begin{equation}
< \mathcal {X},\mathcal {Y} >=\int d^8z_1d^8z_2\overline {\mathcal {X}(z_1)}\mathcal {K}(z_1,z_2)\mathcal {Y}(z_2)
\end{equation}
and the Hilbert space scalar product is
\begin{equation}
(\mathcal {X},\mathcal {Y})=<\mathcal {X},\omega \mathcal {Y}>
\end{equation}
A more interesting Krein structure from the physical point of view is given by (5.3) taking
\begin{equation}
\mathcal {K}=\begin{pmatrix} \frac{1}{16}D^2\bar D^2 K_m & \frac{m}{4}D^2 K_m &0 \\ \frac{m}{4}\bar D^2 K_m& \frac{1}{16}\bar D^2D^2 K_m &0 \\ 0&0& P_T K_0 \end{pmatrix}
\end{equation}
Consequently the Hilbert space structure (5.4) is induced by
\begin{equation}
\omega \mathcal {K}=\begin{pmatrix} \frac{1}{16}D^2\bar D^2 K_m & \frac{m}{4}D^2 K_m &0 \\ \frac{m}{4}\bar D^2 K_m& \frac{1}{16}\bar D^2D^2 K_m &0 \\ 0&0& -P_T K_0 \end{pmatrix}
\end{equation}
For simplicity of the notation in (5.5),(5.6) we have replaced $K_{0m},K_{00}$ by $K_m,K_0 $. This matrix kernel governs the (positive definite) Fock space representation of the matter multiplet together with the  massless supersymmetric vector field providing a concrete realization of this noninteracting system. It provides a convenient rigorous framework for the tentative causal perturbative study of the supersymmetric gauge theory in the sense of Epstein and Glaser \cite{EG}(at least in the abelian case).

\end{document}